\documentclass[traditabstract]{aa} % for the abstract without structuration 

\usepackage{graphicx}
\usepackage{txfonts}

\usepackage[backref,breaklinks,colorlinks,citecolor=blue]{hyperref}        
\usepackage[all]{hypcap}

\def\ms{\,m\,s$^{-1}$}         %m.s -1
\def\kms{\,km\,s$^{-1}$}       %km.s -1
\def\msol{$M_\odot$}		%Msun
\def\rsol{$R_\odot$}		%Rsun
\def\mstar{$M_*$}		%Mstar
\def\rstar{$R_*$}		%Rstar
\def\densstar{$\rho_*$}		%Pstar
\def\arstar{$a/R_*$}		%a/Rstar
\def\mplanet{$M_{\rm P}$}	%Mplanet
\def\rplanet{$R_{\rm P}$}	%Rplanet
\def\mjup{$M_{\rm Jup}$}	%Mjup
\def\rjup{$R_{\rm Jup}$}	%Rjup
\def\msat{$M_{\rm Sat}$}	%Mnep
\def\mnep{$M_{\rm Nep}$}	%Mnep
\def\mearth{$M_{\rm Earth}$}	%Mearth

\def\teff{$T_{\rm eff}$}
\def\feh{[Fe/H]}
\def\logg{$\log g_*$}
\def\vsini{$v_* \sin I_*$}

\def\kms{km\, s$^{-1}$}

\newcommand{\leftcell}[1]{\multicolumn{1}{l}{#1}}

\begin{document}

   \title{The discoveries of WASP-91b, WASP-105b and WASP-107b: two warm Jupiters and a planet in the transition region between ice giants and gas giants\thanks{Based on observations made with: the WASP-South photometric survey instrument, the 0.6-m TRAPPIST robotic imager, and the EulerCam camera and the CORALIE spectrograph mounted on the 1.2-m Euler-Swiss telescope. The photometric time-series and radial-velocity data used in this work are available at the CDS via anonymous ftp to cdsarc.u-strasbg.fr (130.79.128.5) or via http://cdsarc.u-strasbg.fr/viz-bin/qcat?J/A+A/604/A110}.}
   
	\titlerunning{The discoveries of WASP-91b, WASP-105b and WASP-107b}
	\authorrunning{D.~R.~Anderson et al.}

   \author{D.~R.~Anderson\inst{1}
	   \and
	   A.~Collier~Cameron\inst{2}
	   \and
	   L.~Delrez\inst{3,4}
	   \and
	   A.~P.~Doyle\inst{5}
	   \and
	   M.~Gillon\inst{3}
	   \and
	   C.~Hellier\inst{1}
	   \and
	   E.~Jehin\inst{3}
	   \and
	   M.~Lendl\inst{3,6}
	   \and
	   P.~F.~L.~Maxted\inst{1}
	   \and
	   N.~Madhusudhan\inst{4}
	   \and
	   F.~Pepe\inst{6}
	   \and
	   D.~Pollacco\inst{5}
	   \and
	   D.~Queloz\inst{4}
	   \and
	   D.~S\'egransan\inst{6}
	   \and
	   B.~Smalley\inst{1}
	   \and
	   A.~M.~S.~Smith\inst{1,7,8}
	   \and
	   A.~H.~M.~J.~Triaud\inst{6,9,10,11}
	   \and
	   O.~D.~Turner\inst{1}
	   \and
	   S.~Udry\inst{6}
	   \and
	   R.~G.~West\inst{5}
	   }

   \institute{Astrophysics Group, Keele University, Staffordshire ST5 5BG, UK\\
        	  \email{d.r.anderson@keele.ac.uk}
		  \and
		  SUPA, School of Physics and Astronomy, University of St. Andrews, 
       North Haugh, Fife KY16 9SS, UK
	      \and
		  Institut d'Astrophysique et de G\'eophysique,  Universit\'e de 
       Li\`ege,  All\'ee du 6 Ao\^ut, 17,  Bat.  B5C, Li\`ege 1, Belgium
	      \and
		  Cavendish Laboratory, J J Thomson Avenue, Cambridge CB3 0HE, UK
		  \and
		  Department of Physics, University of Warwick, Coventry CV4 7AL, UK
		  \and
		  Observatoire de Gen\`eve, Universit\'e de Gen\`eve, 51 Chemin 
       des Maillettes, 1290 Sauverny, Switzerland
	      \and
		  N. Copernicus Astronomical Centre, Polish Academy of Sciences, Bartycka 18, 00-716, Warsaw, Poland
		  \and
          Institute of Planetary Research, German Aerospace Center, Rutherfordstrasse 2, 12489 Berlin, Germany
          \and
		  Centre for Planetary Sciences, University of Toronto at Scarborough, 1265 Military Trail, Toronto, ON M1C 1A4, Canada
		  \and
		  Department of Astronomy \& Astrophysics, University of Toronto, Toronto, ON M5S 3H4, Canada
		  \and
		  Institute of Astronomy, Madingley Road, CB3 0HA, United Kingdom
			  }

   \date{Received 14 January 2017 / Accepted 11 February 2017}

\abstract{We report the discoveries of three transiting exoplanets. 
WASP-91b is a warm Jupiter (1.34 \mjup, 1.03 \rjup) in a 2.8-day orbit around a metal-rich K3 star. 
WASP-105b is a warm Jupiter (1.8 \mjup, 0.96 \rjup) in a 7.9-day orbit around a metal-rich K2 star. 
WASP-107b is a warm super-Neptune/sub-Saturn (0.12 \mjup, 0.94 \rjup) in a 5.7-day orbit around a solar-metallicity K6 star. Considering that giant planets seem to be more common around stars of higher metallicity and stars of higher mass, it is notable that the hosts are all metal-rich, late-type stars. 
With orbital separations that place both WASP-105b and WASP-107b in the weak-tide regime, measurements of the alignment between the planets' orbital axes and their stars' spin axes may help us to understand the inward migration of short-period, giant planets.\newline
The mass of WASP-107b (2.2 \mnep, 0.40 \msat) places it in the transition region between the ice giants and gas giants of the Solar System. 
Its radius of 0.94 \rjup\ suggests that it is a low-mass gas giant with a H/He-dominated composition. 
The planet thus sets a lower limit of 2.2 \mnep\ on the planetary mass above which large gaseous envelopes can be accreted and retained by proto-planets on their way to becoming gas giants.
We may discover whether WASP-107b more closely resembles an ice giant or a gas giant by measuring its atmospheric metallicity via transmission spectroscopy, for which WASP-107b is a very good target.
}

   \keywords{planetary systems 
             -- stars: individual: WASP-91 
             -- stars: individual: WASP-105 
			 -- stars: individual: WASP-107}

   \maketitle
%
%________________________________________________________________

\section{Introduction}

The observation that the fraction of stars with giant planets increases with both stellar metallicity and mass is suggestive of planetary formation by core accretion (e.g. \citealt{2004A&A...415.1153S, 2010PASP..122..905J}). 
Under the core accretion model (e.g. \citealt{1996Icar..124...62P}), a gas giant results when planetesimals coagulate to form a rocky core, which then accretes a gaseous envelope. 
The Solar System's gas giants, Jupiter and Saturn (0.30 \mjup, 0.84 \rjup), are more than 90\,\% H/He by mass, which contrasts with the figure of 20\,\% for the less massive ($\sim$0.05 \mjup) and smaller ($\sim$0.35 \rjup) ice giants, Neptune and Uranus \citep{2005AREPS..33..493G}. 
One challenge faced by models attempting to explain the formation of Neptune and Uranus is to avoid the runaway gas accretion that otherwise would have turned the planets into gas giants (e.g. \citealt{2014ApJ...789...69H}).

Giant planets in few-day orbits, or `warm/hot Jupiters', are thought to have formed farther out and then migrated inwards via interaction with the gas disc or via a high-eccentricity pathway \citep{1996Natur.380..606L,1996Sci...274..954R}. 
Planet-disc migration is expected to preserve alignment between the stellar spin and planetary orbital axes (e.g. \citealt{2009ApJ...705.1575M}), whereas high-eccentricity migration is expected to produce a broad range of misalignments (e.g. \citealt{2007ApJ...669.1298F}). 
The ensemble of available measurements has been interpreted as evidence that hot Jupiters arise via high-eccentricity migration \citep{2010ApJ...718L.145W}, though planet-disc migration is likely to play a role (e.g. \citealt{2015ApJ...800L...9A}).

In this paper, we present the discoveries of three transiting exoplanets by the WASP survey: WASP-91b and WASP-105b are warm Jupiters orbiting metal-rich, early/mid-K stars; and WASP-107b is a warm super-Neptune/sub-Saturn orbiting a solar-metallicity, late-K star. 

\section{Observations}
\label{sec:obs}
WASP-South images one third of the visible South-African sky (avoiding the galactic plane and the south pole) every $\sim$10 minutes and is sensitive to the detection of giant planets transiting bright stars ($V$ = 9--13). 
The survey and the search techniques are described in 
\citet{2006PASP..118.1407P} and \citet{2006MNRAS.373..799C, 2007MNRAS.380.1230C}. 

We routinely investigate the promising transit signals that we find in WASP lightcurves with the EulerCam imager and the CORALIE spectrograph, both of which are mounted on the 1.2-m Euler-Swiss telescope, and the 0.6-m TRAPPIST imager and \citep{2012A&A...544A..72L, 2000A&A...354...99Q, 2011A&A...533A..88G, 2011Msngr.145....2J}. 
We provide a summary of our observations of the three target stars in \autoref{tab:obs}. 
TRAPPIST performed meridian flips at the following times (BJD$-$2\,450\,000): 6560.756 (WASP-105 transit of 2013 September 24); 7001.546 (WASP-105 transit of 2014 December 9); 6428.575 (WASP-107 transit of 2013 May 15); and 6754.683 (WASP-107 transit of 2014 April 6). We partioned the resulting lightcurves prior to fitting to allow for flux offsets. 
We interpret the bump in the lightcurve of WASP-107 around mid-transit on 2013 Feb 18 as having been caused by the planet occulting a star spot. 

\begin{figure}
\includegraphics[width=84mm]{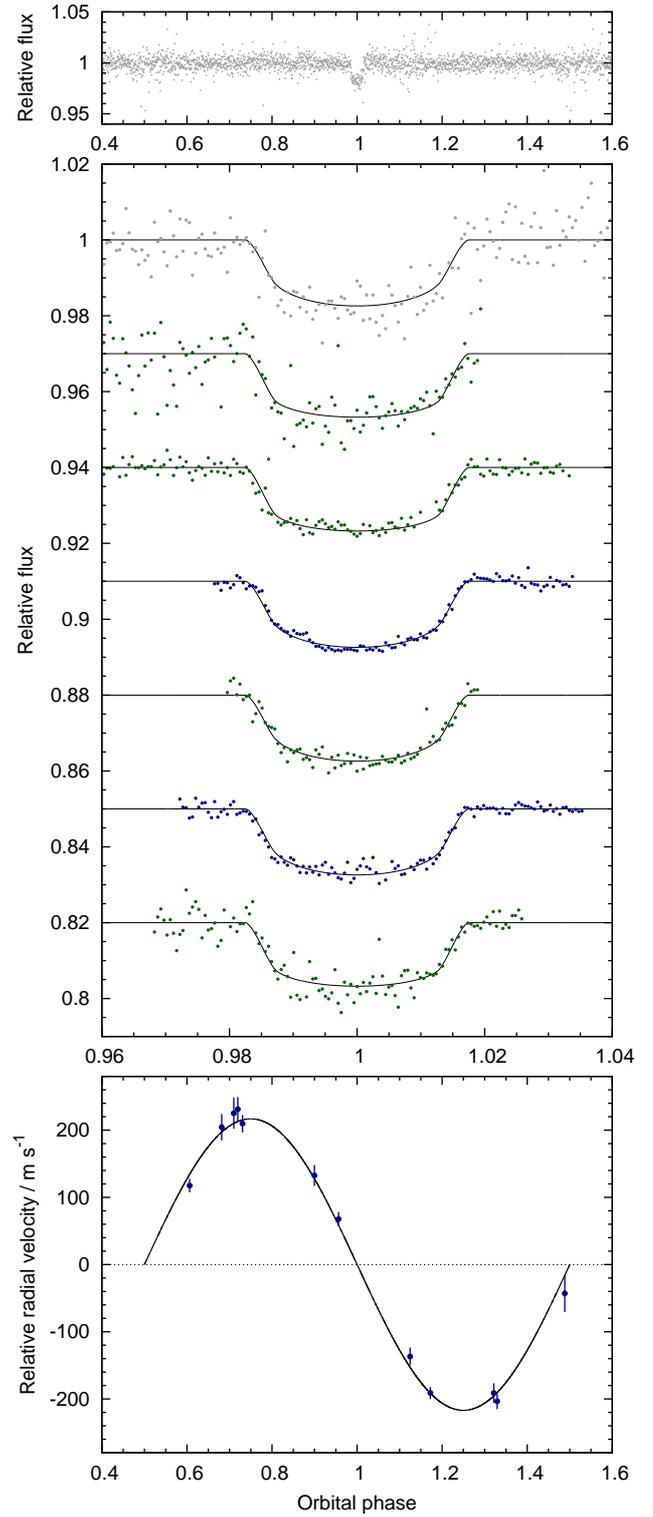}
\caption{WASP-91b discovery data. 
{\it Top panel}: WASP lightcurve folded on the transit ephemeris. 
{\it Middle panel}: Transit lightcurves from WASP (grey), TRAPPIST (green) and EulerCam (blue), offset for clarity, binned with a bin width of two minutes, and plotted chronologically with the most recent at the bottom. 
The best-fitting transit model is superimposed. 
{\it Bottom panel}: The CORALIE radial velocities with the best-fitting circular orbital model. 
\label{fig:w91-rv-phot}}
\end{figure}

\begin{figure}
\includegraphics[width=84mm]{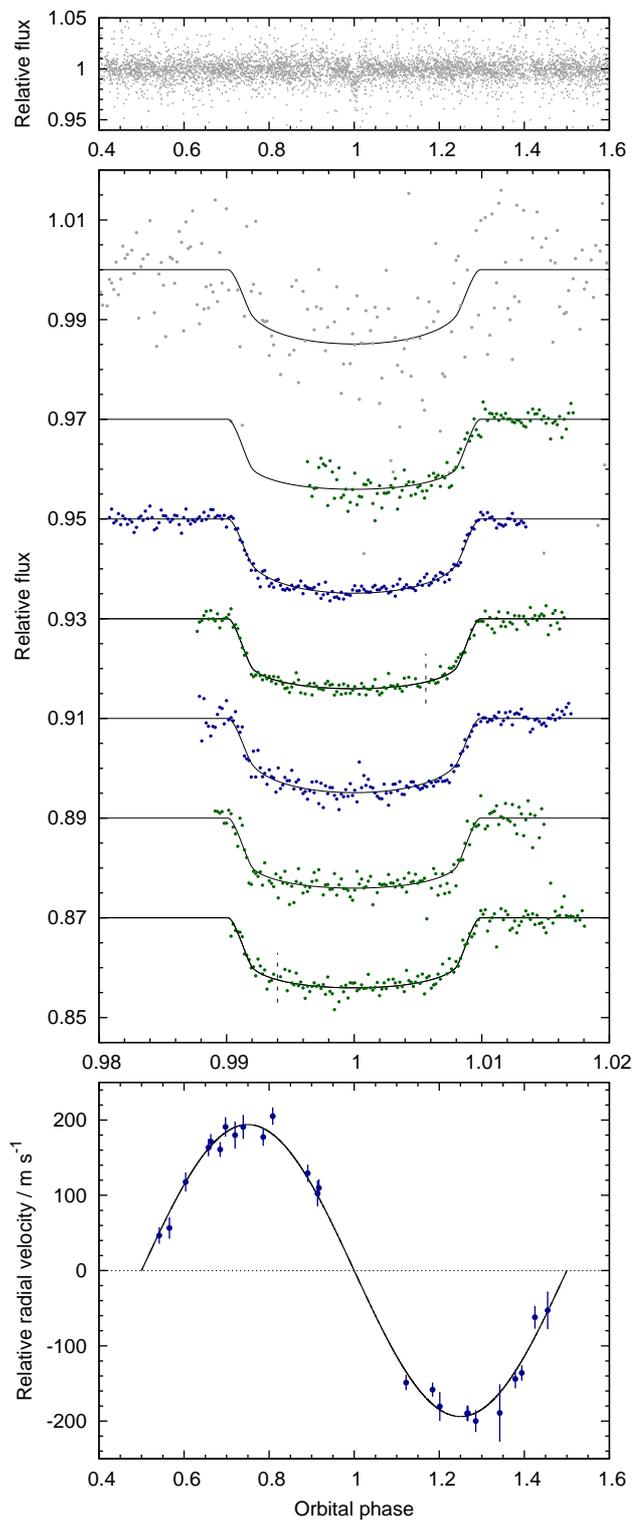}
\caption{WASP-105b discovery data. Caption as for \autoref{fig:w91-rv-phot}. Data partioning due to TRAPPIST's meridian flips are indicated by vertical dashed lines.
\label{fig:w105-rv-phot}}
\end{figure}

\begin{figure}
\includegraphics[width=84mm]{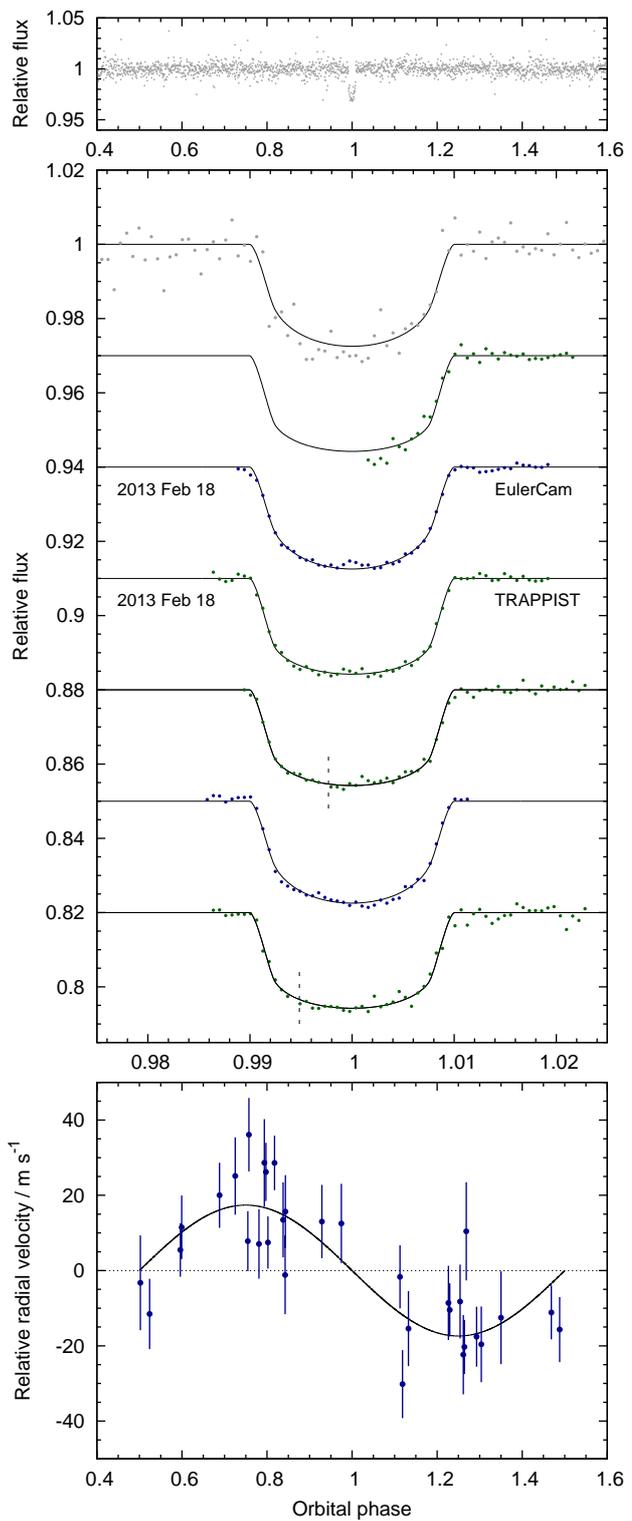}
\caption{WASP-107b discovery data. Caption as for \autoref{fig:w91-rv-phot}. 
The planet appears to have passed over a star spot during the transit of 2013 Feb 18, with a bump more evident in the lightcurve from EulerCam (Gunn $r$) than TRAPPIST (Sloan $z$). 
The difference is expected due to the bluer passband employed by EulerCam, in which the contrast between star spots and the surrounding photosphere will be greater. Also, the diameter of the TRAPPIST telescope is half that of the Euler telescope (0.6 m versus 1.2 m), so the data are noisier.
Data partioning due to TRAPPIST's meridian flips are indicated by vertical dashed lines. \label{fig:w107-rv-phot}}
\end{figure}

\begin{figure}
\includegraphics[]{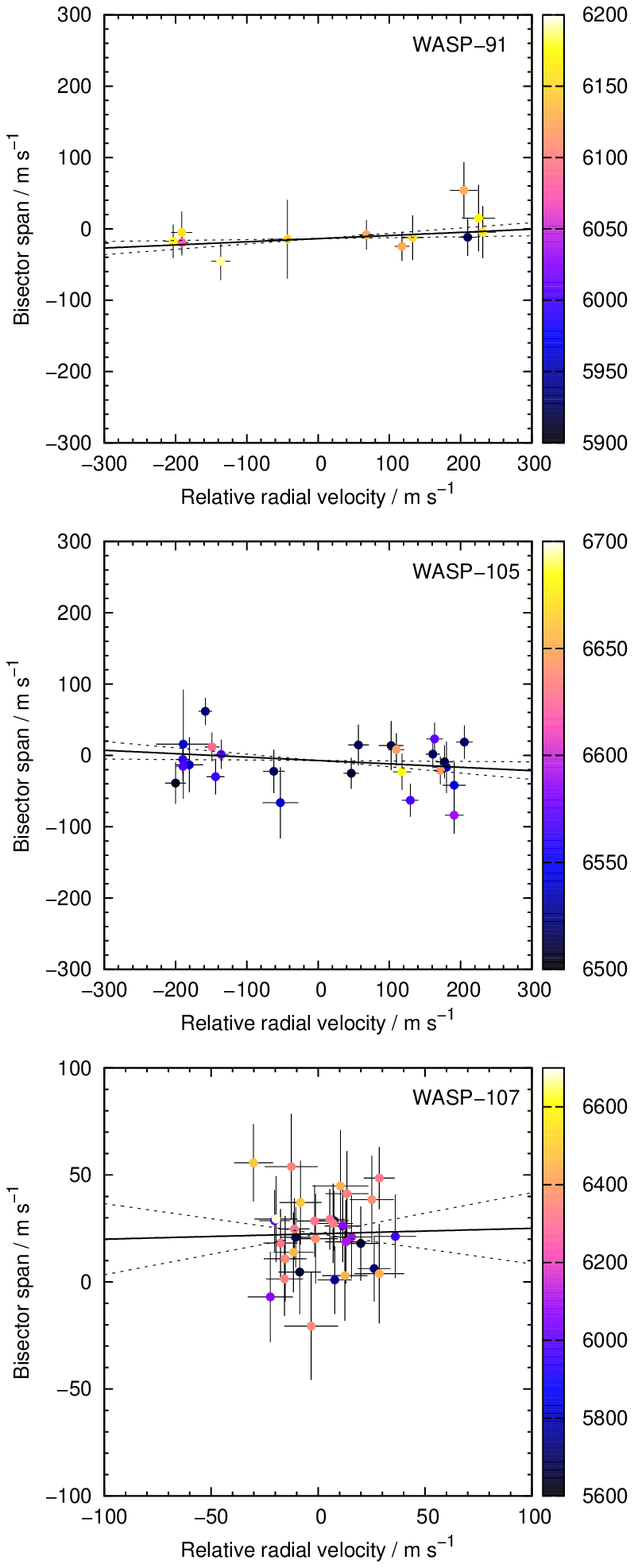}
\caption{The absence of correlation between bisector span and radial velocity for the three stars excludes transit mimics. 
The solid line is the best linear fit to the data and the dotted lines are the 1-$\sigma$ limits on the gradient. 
The Julian date of the observation (BJD $-$ 2\,450\,000) is represented by the symbol colour.
\label{fig:bis}} 
\end{figure}

The radial-velocity (RV) measurements that we computed from the CORALIE spectra exhibit variations with similar periods as the photometric dimmings seen in the WASP lightcurves and with amplitudes consistent with planetary-mass companions. 
The photometry and RVs are plotted for each system in Figures~\ref{fig:w91-rv-phot}, \ref{fig:w105-rv-phot} and \ref{fig:w107-rv-phot}.
The absence of a significant correlation between bisector span and RV supports our conclusion that the observed periodic dimmings and RV variations are caused by transiting planets (\autoref{fig:bis}). 

\begin{table}
\scriptsize
\centering
\caption{Summary of observations}
\label{tab:obs}
\begin{tabular}{lcrcl}
\hline
\hline
\leftcell{Facility} & \leftcell{Date} & \leftcell{$N_{\rm obs}$} & \leftcell{$T_{\rm exp}$ [s]} & \leftcell{Filter} \\
\hline
{\bf WASP-91:}\\
WASP-South	    & 2010 Jun--2011 Dec	& 16\,800& 30 & Broad (400--700 nm)	\\
Euler/CORALIE	& 2012 Jan--Sep	& 12	& 1800		& Spectroscopy \\
TRAPPIST	& 2012 Jul 20			& 578	& 15 & $I$+$z'$			\\
TRAPPIST	& 2012 Aug 31			& 934	& 12 & $I$+$z'$			\\
Euler/EulerCam	& 2012 Oct 12			& 172	& 70 & Gunn-$r$ \\
TRAPPIST	& 2012 Oct 29			& 395	& 15 & $R$			\\
Euler/EulerCam	& 2013 Jun 04			& 131	& 110 & Gunn-$r$ \\
TRAPPIST	& 2013 Jun 04			& 794	& 8 & $I$+$z'$			\\
{\bf WASP-105:}\\
WASP-South	    & 2010 Jun--2011 Dec	& 11\,200& 30 & Broad (400--700 nm)	\\
Euler/CORALIE	& 2013 Aug--2014 Jan	& 25	& 1800		& Spectroscopy \\
TRAPPIST	& 2013 Jul 23			& 703	& 10 & $I$+$z'$			\\
Euler/EulerCam	& 2013 Sep 24			& 288	& 70 & Gunn-$r$ \\
TRAPPIST	& 2013 Sep 24			& 906	& 12 & $I$+$z'$			\\
Euler/EulerCam	& 2013 Oct 02			& 257	& 70 & Gunn-$r$ \\
TRAPPIST	& 2013 Nov 26			& 792	& 12 & $I$+$z'$			\\
TRAPPIST	& 2014 Dec 09			& 1072	& 7 & $I$+$z'$			\\
{\bf WASP-107:}\\
WASP-South	    & 2009 Feb--2010 Jun	& 9\,350& 30 & Broad (400--700 nm)	\\
Euler/CORALIE	& 2011 Mar--2014 Jan	& 32	& 1800		& Spectroscopy \\
TRAPPIST	& 2013 Jan 09			& 532	& 10 & $z'$			\\
Euler/EulerCam	& 2013 Feb 18		& 264	& 50 & Gunn-$r$ \\
TRAPPIST	& 2013 Feb 18			& 864	& 10 & $z'$			\\
TRAPPIST	& 2013 May 15			& 814	& 10 & $z'$			\\
Euler/EulerCam	& 2014 Feb 02		& 154	& 75 & Gunn-$r$ \\
TRAPPIST	& 2014 Apr 06			& 692	& 10 & $z'$			\\
\hline
\end{tabular}
\end{table}

\section{Stellar parameters from spectra}
\label{sec:stars}
The individual CORALIE spectra were co-added after correcting for the orbital motion of the star, giving average S/N ratios of 55:1, 80:1 and 120:1 for WASP-91, WASP-105 and WASP-107, respectively. We performed the spectral analysis using the procedures detailed in \cite{2013MNRAS.428.3164D}. For each star the effective temperature ($T_{\rm eff}$) was obtained using the H$\alpha$ line and surface gravity (\logg) was determined from the Na~D and Mg~b lines. Iron abundances were obtained from the analysis of equivalent width measurements of several unblended Fe~{\sc i} lines.

Projected equatorial rotation velocities (\vsini) were determined by fitting the profiles of the Fe~{\sc i} lines after convolving with the instrumental resolution ($R$ = 55\,000).
For WASP-105 a macroturbulent velocity of 0.9 $\pm$ 0.3 {\kms} was adopted from \citep{2008oasp.book.....G}. In the cases of WASP-91 and WASP-107, macroturbulence was assumed to be zero, since for mid/late K stars it is expected to be lower than that of thermal broadening \citep{2008oasp.book.....G}.
The results are given in the top panel of \autoref{tab:mcmc}.

\section{Stellar rotation from lightcurve modulation}

\begin{figure}
\mbox{\includegraphics[width=0.49\textwidth]{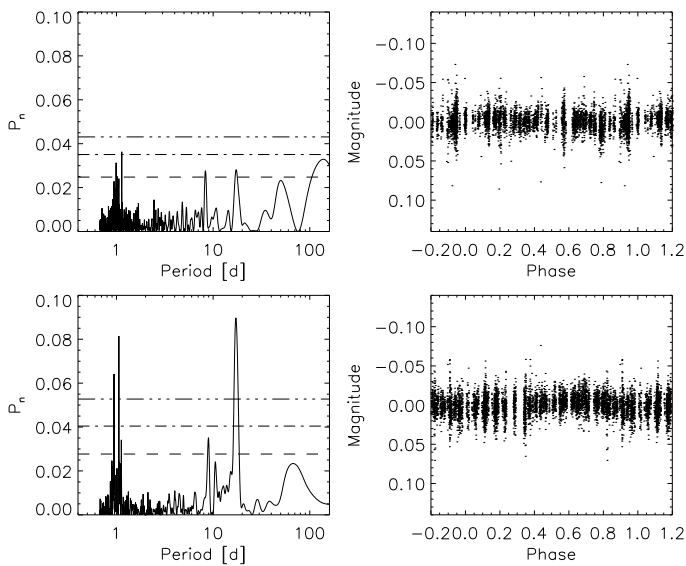}}
\caption{Left: Periodograms of the WASP lightcurves for WASP-107 obtained
during 2009 (upper panel) and 2010 (lower panel). Horizontal lines indicate
false-alarm probability levels 0.1, 0.01 and 0.001. Right: Lightcurves folded
on the assumed rotation period of 17.1\,days for data obtained during 2009
(upper panel) and 2010 (lower panel).
\label{ProtFig}}
\end{figure}

 The WASP light curves of WASP-107 show a periodic modulation with an amplitude
of about 0.4 per cent and a period of about 17 days. We assume this is due to
the combination of the star's rotation and magnetic activity, i.e., star
spots. We used the sine-wave fitting method described in
\citet{2011PASP..123..547M} to refine this estimate of the amplitude and
period of the modulation. 
Variability due to star spots is not expected to be
coherent on long timescales as a consequence of the finite lifetime of
star-spots and differential rotation in the photosphere so we analysed the two
seasons of data for WASP-107 separately.  We removed the transit signal from
the data prior to calculating the periodograms by subtracting a simple transit
model from the lightcurve and also removed low-frequency noise by subtracting
a straight line fit by least-squares to the data from each season. We
calculated periodograms over 8192 uniformly spaced frequencies from 0 to 1.5
cycles/day. The false alarm probability (FAP) is calculated using a boot-strap
Monte Carlo method also described in \citet{2011PASP..123..547M}. The results
are given in \autoref{ProtTable} and the periodograms and lightcurves 
are shown in \autoref{ProtFig}. There is a clear signal at a period of
$P$ = 17.1\,d in the 2010 season of data that is also in the data from 2009 at $P$ = 17.3\,d, though at a lower significance. The periodogram of the data
from 2009 also shows a peak at $P$ = 8.3\,d, which we assume is the second-harmomic
of the rotation period due of multiple spot groups on surface of the star during
this observing season. The strongest peak in this periodogram at 1.134\,d can
then be ascribed to the 1-day alias of this second-harmomic.

Assuming this to be the case, we obtain a value for the rotation period of
$P_{\rm rot} = 17  \pm 1$\,d, where the error in this value is taken from the
full-width at half-maximum of the peak in the periodogram of the data from
2010.  This rotation period together with our
estimate of the stellar radius (\autoref{sec:mcmc}) implies a value for the stellar rotation velocity of 
$v_*$ = $2.0 \pm 0.1$\,\kms. 
This compares well with the spectroscopic estimate of the projected equatorial rotation velocity of \vsini\ = $2.5 \pm 0.8$ \kms. We used a least-squares
fit of a sinusoidal function and its first harmonic to model the rotational
modulation in the lightcurves for each camera and season with the rotation
period fixed at $P_{\rm rot}  = 17.1$\,d. We then subtracted this harmonic
series fit from the original lightcurve prior to our analysis of the transit (\autoref{sec:mcmc}).

For WASP-91 and WASP-105 a similar analysis lead to upper limits 
with 95\,per~cent confidence of 0.8\,mmag and 0.7\,mmag  for
the amplitude of any sinusoidal signal over the same frequency range.

\clearpage

%%%%%%%%%%%%%%%%%%%%%%%
%MCMC paramters 
%%%%%%%%%%%%%%%%%%%%%%%
\begin{table*} 
\caption{System parameters} 
\label{tab:mcmc}
\small
\begin{tabular}{lcccc}
\hline
\hline
Parameter & Symbol (Unit) & WASP-91 & WASP-105 & WASP-107\\ 
\hline 
\multicolumn{5}{l}{Stellar parameters, including from the spectra:}\\
Constellation	& & Tucana & Phoenix & Virgo \\
Right Ascension	& & $\rm 23^{h} 51^{m} 22\fs89$		& $\rm 01^{h} 36^{m} 40\fs24$		& $\rm 12^{h} 33^{m} 32\fs84$		\\
Declination		& & $\rm -70\degr 09\arcmin 10\fs2$	& $\rm -50\degr 39\arcmin 32\fs5$	& $\rm -10\degr 08\arcmin 46\fs1$	\\	
$V_{\rm mag}$	& & 12.0	& 12.1	& 11.6	\\
$K_{\rm mag}$	& & 9.7		& 9.9	& 8.6	\\
Spectral type\tablefootmark{a}   & &  K3 & K2 & K6 \\
Stellar effective temperature & \teff\ (K) &   4920 $\pm$ 80 &       5070 $\pm$ 130        &       4430 $\pm$ 120 \\
Stellar surface gravity & \logg\ (cgs)   &   4.3 $\pm$ 0.2 & 4.2 $\pm$ 0.2        &          4.5 $\pm$ 0.1 \\       
Projected equatorial rotation velocity & \vsini\  / \kms   &   2.4 $\pm$ 0.4 &  1.7 $\pm$         1.9 &            2.5 $\pm$ 0.8 \\ 
Stellar metallicity\tablefootmark{b} & \feh   &   +0.19 $\pm$ 0.13 & +0.28 $\pm$ 0.16             &      +0.02$\pm$ 0.10 \\      
Lithium abundance & $\log A$(Li) &   $<$ 0.5 &      $<$ 0.2                      & $<$ $-$0.3 \\
%Mass / \msol     &   0.89 $\pm$ 0.08  & 0.99 $\pm$0.11 & 0.68 $\pm$ 0.06 \\
%Radius / \rsol  &   1.13 $\pm$ 0.27 & 1.28 $\pm$ 0.34 & 0.75 $\pm$ 0.10 \\
\hline
\multicolumn{5}{l}{MCMC proposal parameters:}\\
Orbital period & $P$ (d) & 2.798581 $\pm$ 0.000003 & 7.87288 $\pm$ 0.00001 & 5.721490 $\pm$ 0.000002 \\
Epoch of mid-transit & $T_{\rm c}$ (BJD) & 2\,456\,297.7190 $\pm$ 0.0002 & 2\,456\,600.0765 $\pm$ 0.0002 & 2\,456\,514.4106 $\pm$ 0.0001 \\
Transit duration & $T_{\rm 14}$ (d) & 0.0976 $\pm$ 0.0008 & 0.1550 $\pm$ 0.0006 & 0.1147 $\pm$ 0.0003 \\
Planet-to-star area ratio & $R_{\rm P}^{2}$/R$_{*}^{2}$ & 0.0150 $\pm$ 0.0003 & 0.0120 $\pm$ 0.0001 & 0.0217 $\pm$ 0.0002 \\
Impact parameter\tablefootmark{c} & $b$ & 0.51 $\pm$ 0.04 & 0.10 $\pm$ 0.08 & 0.09 $\pm$ 0.07 \\
Stellar reflex velocity semi-amplitude & $K_{\rm 1}$ (m s$^{-1}$) & 217 $\pm$ 5 & 194 $\pm$ 3 & 17 $\pm$ 2 \\
Systemic velocity & $\gamma$ (m s$^{-1}$) & 2\,782 $\pm$ 4 & 24\,676 $\pm$ 2 & 14\,160 $\pm$ 2 \\
Orbital eccentricity & $e$ & 0 (adopted; $<$ 0.07  at 2\,$\sigma$) & 0 (adopted; $<$ 0.04 at 2\,$\sigma$) & 0 (adopted; $<$ 0.4  at 2\,$\sigma$) \\
\hline
\multicolumn{5}{l}{MCMC derived parameters:}\\
Scaled orbital separation & $a$/\rstar & $9.1 \pm 0.3$ & $17.9 \pm 0.2$ & $18.2 \pm 0.1$ \\
Orbital inclination & $i$ ($^\circ$) & 86.8 $\pm$ 0.4 & 89.7 $\pm$ 0.2 & 89.7 $\pm$ 0.2 \\
Transit ingress/egress duration & $T_{\rm 12}=T_{\rm 34}$ (d) & 0.0139 $\pm$ 0.0009 & 0.0154 $\pm$ 0.0003 & 0.0148 $\pm$ 0.0002 \\
Stellar mass & $M_{\rm *}$ ($M_{\rm \odot}$) & 0.84 $\pm$ 0.07 & 0.89 $\pm$ 0.09 & 0.69 $\pm$ 0.05 \\
Stellar radius & $R_{\rm *}$ ($R_{\rm \odot}$) & 0.86 $\pm$ 0.03 & 0.90 $\pm$ 0.03 & 0.66 $\pm$ 0.02 \\
Stellar surface gravity & $\log g_{*}$ (cgs) & 4.49 $\pm$ 0.03 & 4.48 $\pm$ 0.02 & 4.64 $\pm$ 0.01 \\
Stellar density & $\rho_{\rm *}$ ($\rho_{\rm \odot}$) & 1.3 $\pm$ 0.1 & 1.23 $\pm$ 0.03 & 2.45 $\pm$ 0.05 \\
Planetary mass & $M_{\rm P}$ ($M_{\rm Jup}$) & 1.34 $\pm$ 0.08 & 1.8 $\pm$ 0.1 & 0.12 $\pm$ 0.01 \\
Planetary radius & $R_{\rm P}$ ($R_{\rm Jup}$) & 1.03 $\pm$ 0.04 & 0.96 $\pm$ 0.03 & 0.94 $\pm$ 0.02 \\
Planetary surface gravity & $\log g_{\rm P}$ (cgs) & 3.46 $\pm$ 0.03 & 3.64 $\pm$ 0.01 & 2.49 $\pm$ 0.05 \\
Planetary density & $\rho_{\rm P}$ ($\rho_{\rm J}$) & 1.2 $\pm$ 0.1 & 2.0 $\pm$ 0.1 & 0.14 $\pm$ 0.02 \\
Orbital major semi-axis & $a$ (au)  & 0.037 $\pm$ 0.001 & 0.075 $\pm$ 0.003 & 0.055 $\pm$ 0.001 \\
Planetary equil. temperature\tablefootmark{d} & $T_{\rm eql}$ (K) & 1160 $\pm$ 30 & 900 $\pm$ 20 & 770 $\pm$ 60 \\
\hline 
\end{tabular}
\tablefoot{
\tablefoottext{a}{Iron abundances are relative to the solar values of \cite{2009ARA&A..47..481A}.}
\tablefoottext{b}{Spectral type estimated from \teff\ using the table in \cite{2008oasp.book.....G}.}
\tablefoottext{c}{Impact parameter is the distance between the centre of the stellar disc and the transit chord: $b = a \cos i / R_{\rm *}$.}
\tablefoottext{d}{Equilibrium temperature calculated assuming zero albedo and efficient redistribution of heat from the planet's presumed permanent day-side to its night-side.}
}
\end{table*}

\begin{table}
 \caption{Periodogram analysis of the WASP lightcurves for WASP-107. 
\label{ProtTable}}
 \begin{tabular}{@{}lrrrrr}
\hline
\hline
  \multicolumn{1}{@{}l}{Season} &
  \multicolumn{1}{l}{Dates} &
  \multicolumn{1}{l}{$N$} &
  \multicolumn{1}{l}{$P$ [d]} &
  \multicolumn{1}{l}{$a$ [mmag]} &
  \multicolumn{1}{l}{FAP}\\
 \noalign{\smallskip}
\hline
2009& 4867\,--\,5010 &  4029   & 1.134  &  0.003  & 0.056  \\
2010& 5233\,--\,5376 &  5315   & 17.17  &  0.004  &$<0.001$\\
 \noalign{\smallskip}
\hline
 \end{tabular}   
\tablefoot{
Observing dates are JD $-$ 2\,450\,000,  $N$ is the number of observations used in the analysis,
$a$ is the semi-amplitude of the best-fit sine wave at the period $P$ found
in the periodogram with false-alarm probability FAP.
}
 \end{table}     

\section{System parameters from the RV and transit data}
\label{sec:mcmc}

We determined the parameters of each system by fitting the photometric and radial-velocity data simultaneously 
using the current version of the 
Markov-chain Monte Carlo (MCMC) code described by \citet{2007MNRAS.380.1230C} 
and \citet{2015A&A...575A..61A}.
The transit lightcurves were modelled using the formulation of 
\citet{2002ApJ...580L.171M} and limb-darkening was accounted for using the four-parameter non-linear law of \citet{2000A&A...363.1081C, 2004A&A...428.1001C}. 

Stellar density is measured from the transit lightcurves, but we require a constraint on stellar mass for a full characterisation of the system. For that we used the {\sc bagemass} stellar evolution MCMC code of \citet{2015A&A...575A..36M}, using the values of 
\densstar\ from an initial MCMC run and the values of \teff\ and \feh\ from the spectral analysis. From {\sc bagemass} we obtained values of stellar mass (\mstar) of 0.840 $\pm$ 0.032 \msol, 0.891 $\pm$ 0.047 \msol\ and 0.691 $\pm$ 0.025 \msol\ for WASP-91, WASP-105 and WASP-107, respectively. In our final MCMC analyses, we drew a value of \mstar\ at each MCMC step from a normal distribution with mean and standard deviation equal to the {\sc bagemass}-derived values, but with an error bar larger by a factor 2 to allow for uncertainties due to the unknown helium abundances and the effects of magnetic activity on the mass-radius relation.

In initial MCMC runs we modelled eccentric orbits, but for no system do we find compelling evidence of a non-circular orbit. We thus adopt circular orbits, which \citet{2012MNRAS.422.1988A} argue is the prudent choice for short-period, $\sim$Jupiter-mass planets in the absence of evidence to the contrary. We place 2-$\sigma$ upper limits on orbital eccentricity of 0.07, 0.04 and 0.4 for WASP-91b, WASP-105b and WASP-107b, respectively. 

We present the system parameters from our final MCMC analyses in \autoref{tab:mcmc} and we plot the best fits to the radial-velocity data and the photometric data in 
Figures~\ref{fig:w91-rv-phot}, \ref{fig:w105-rv-phot} and \ref{fig:w107-rv-phot}.

\section{Discussion}
\label{sec:discuss}

WASP-91b, a 1.34-\mjup\ planet in a 2.8-day orbit around a K3 star, is the southern-most transiting planet known. 
WASP-105b is a 1.8-\mjup\ planet in a 7.9-day orbit around a K2 star. 
Finally, WASP-107b is a 0.12-\mjup\ planet in a 5.7-day orbit around a K6 star. 
Together with WASP-139b (\mplanet\ = $0.12 \pm 0.02$ \mjup; \citealt{2016arXiv160404195H}), WASP-107b 
is the lowest-mass planet discovered by WASP to date; the next lowest are WASP-29b (0.24 \mjup; \citealt{2010ApJ...723L..60H}) and WASP-69b (0.26 \mjup; \citealt{2014MNRAS.445.1114A}). 
Giant planets seem to be more common around both stars of higher metallicity and stars of higher mass (e.g. \citealt{2004A&A...415.1153S, 2010PASP..122..905J}). 
It is interesting to note that all three hosts are K stars and that WASP-91 and WASP-105 are metal rich, whilst the super-Neptune host, WASP-107, is solar metallicity.

WASP-91b (1.03 \rjup) and WASP-105b (0.96 \rjup) are notable as having radii towards the lower end of the envelope for hot Jupiters (\autoref{fig:mass-radius}), though they are as expected from the empirical relation of \citet{2012A&A...540A..99E} based on their semi-major axes and relatively low equilibrium temperatures. 
WASP-107b occupies a sparsely populated region in the planetary mass-radius diagram, with a mass 2.2 times that of Neptune and 0.40 times that of Saturn 
(\autoref{fig:mass-radius}). 
The planet's radius is toward the upper end of the super-Neptune/sub-Saturn regime and it is higher than expected from the empirical relation of \citet{2012A&A...540A..99E} by around 0.30 \rjup, perhaps suggestive of a low-metallicity composition. 

\subsection{WASP-107b and the transition between ice giants and gas giants}
Under the core accretion model of planet formation, planetesimals coagulate to form a rocky core, which rapidly accretes a gaseous envelope once a critical mass of $\sim$10\,\mearth\ is reached \citep{1978PThPh..60..699M}.
One challenge for planet formation models is to explain why ice giants did not become gas giants.

\citet{2014A&A...572A..35L} suggested core growth via the accretion of pebbles, rather than planetesimals, as a solution. 
Under this hypothesis, beyond a threshold mass, a core can halt the accretion of pebbles by gravitationally perturbing the surrounding disc. 
The gas envelope surrounding the core is then no longer supported by accretion heat and so rapidly collapses, resulting in a gas giant. 
Ice giants do not reach this threshold mass, which depends on orbital distance due to the steep increase in the gas scale height in flaring discs. 
This hypothesis offers a neat explanation for the bifurcation of the giants of the Solar System and it can be tested as it predicts both that ice giants in wide orbits are common relative to gas giants and that those gas giants are enriched (core mass $>$ 50 $M_{\rm \oplus}$). 

\begin{figure}
\includegraphics[width=0.49\textwidth]{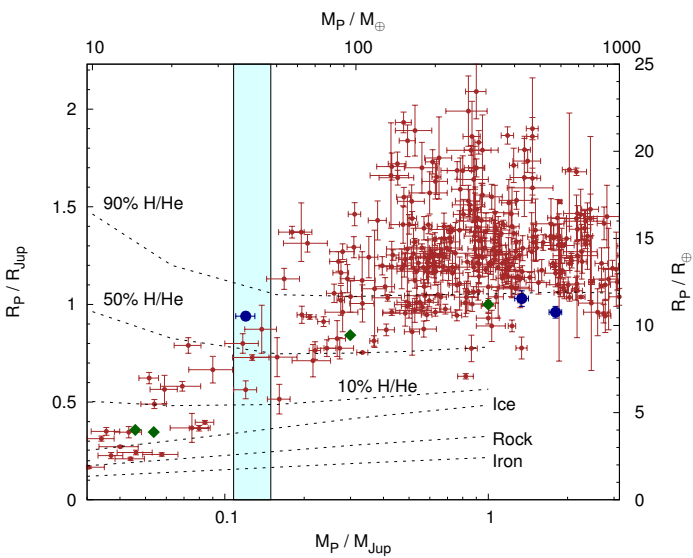}
\caption{Planetary mass-radius diagram showing the planets presented herein (blue circles), the Solar System giants (green diamonds) and the transiting exoplanets (red circles; data from TEPCat; masses measured by the radial-velocity technique to better than 20\% precision). 
The dotted lines depict model planets of pure iron, rock and ice (from \citealt{2007ApJ...659.1661F}) and 3-Gyr isochrones for model planets with various H/He fractions that are irradiated by a Sun-like star at 0.045 AU (from \citealt{2008A&A...482..315B}). 
The cyan rectangle indicates the notional transition region between ice giants and gas giants, where planetary mass is between twice that of Neptune and half that of Saturn. 
\label{fig:mass-radius}}
\end{figure}

WASP-107b has a mass 2.2 times that of Neptune and 0.40 times that of Saturn, but a radius 0.94 times that of Jupiter. This suggests that WASP-107b is a low-mass gas giant, with a H/He-dominated composition (\autoref{fig:mass-radius}). 
We define a notional transition region between ice giants and gas giants which spans a planetary mass of between twice that of Neptune and half that of Saturn (0.11 \mjup\ $<$ \mplanet\ $<$ 0.15 \mjup). 
We know of five planets\footnote{
  We excluded Kepler-9b and Kepler-9c as there is considerable uncertainty regarding their masses, which were inferred from transit-timing variations \citep{2010Sci...330...51H, 2014A&A...571A..38B, 2014ApJ...787...80H}.
} 
with masses in that region: WASP-107b, WASP-139b \citep{2016arXiv160404195H}, HATS-7b \citep{2015ApJ...813..111B}, HATS-8b \citep{2015AJ....150...49B}, and the circumbinary planet Kepler-35b \citep{2012Natur.481..475W}. 
We may be able to discern whether these planets more closely resemble ice giants or gas giants by measuring their atmospheric metallicities, which is achievable by measuring their atmospheric H$_2$O abundances via transmission spectroscopy with HST (e.g. \citealt{2014ApJ...793L..27K}).
The atmospheric metallicities of Neptune and Uranus, the Solar System's ice giants, are far higher than that of Jupiter and Saturn, the Solar System's gas giants: with C/H $\approx$ 80 times the proto-solar abundance as compared to 4--10 times \citep{2014arXiv1405.3752G}.
By measuring the atmospheric metallicity of planets in the transition region, we may gain insight into the planetesimal-accretion history of the planet and better understand both the formation pathways of ice giants and gas giants and the transition from one class to the other. 
For each planet in the transition region, we calculated their predicted atmospheric transmission signal (i.e. the product of the star's $K$-band flux and the area ratio of the planetary atmosphere's annulus to the stellar disc; \autoref{tab:transm}).
This suggests that WASP-107b is the most favourable target for transmission spectroscopy in the transmission region by an order of magnitude. Further, it is predicted to be an order of magnitude more favourable than WASP-43b, whose H$_2$O abundance was measured recently \citep{2014ApJ...793L..27K}.

\begin{table*}
 \caption{Planets in the transmission region between ice giants and gas giants. 
\label{tab:transm}}
 \begin{tabular}{@{}lrrrrrrrrr}
\hline
\hline
Planet & 
\mplanet &
\rplanet & 
$P$ & 
\mstar & 
\rstar & 
\teff & 
$V$ & 
$K$ & 
Transm.
\\

 &
(\mjup) &
(\rjup) & 
(d) &
(\msol) &
(\rsol) &
(K) &
&
&
signal\\
 \noalign{\smallskip}
\hline
WASP-107b 	& 0.12 $\pm$ 0.01 &  0.94 $\pm$ 0.02 &  5.72 &  0.69 $\pm$ 0.05 & 0.66 $\pm$ 0.02 & 4430 $\pm$ 120 &  11.6 &  8.6 & 1000\\
WASP-139b 	& 0.12 $\pm$ 0.02 &  0.80 $\pm$ 0.05 &  5.92 &  0.92 $\pm$ 0.10 & 0.80 $\pm$ 0.04 & 5300 $\pm$ 100 &  12.4 &  10.5 & 94\\
HATS-7b 	& 0.12 $\pm$ 0.01 &  0.56 $\pm$ 0.04 &  3.19 &  0.85 $\pm$ 0.03 & 0.82 $\pm$ 0.04 & 4985 $\pm$ 50  &  13.3 & 11.0 & 21\\
HATS-8b 	& 0.14 $\pm$ 0.02 &  0.87$^{+0.12}_{-0.08}$ &  3.58 &  1.06 $\pm$ 0.04 & 1.09$^{+0.15}_{-0.05}$ & 5679 $\pm$ 50 &  14.0 & 12.7 & 10\\
Kepler-35b	& 0.13 $\pm$ 0.02	& 0.73 $\pm$ 0.01	& 131.46	& 0.888 $\pm$ 0.005	& 1.028 $\pm$ 0.002 & 5606 $\pm$ 150	& 15.9 &	13.9	& 1\\
WASP-43b$^{\dagger}$	& 2.03 $\pm$ 0.05	& $1.04 \pm 0.02$	& 0.81	& 0.72 $\pm$ 0.03	& 0.67 $\pm$ 0.01	& 4520 $\pm$ 120	& 12.4	& 9.3	& 74\\
\hline
 \end{tabular}   
\tablefoot{The data for WASP-107b are from this paper and the data for the other systems were taken from TEPCat. 
The transmission signal (final column) is the product of the star's $K$-band flux and the area ratio of the 
planetary atmosphere's annulus to the stellar disc; the transmission signal values were normalised such that the predicted signal for WASP-107b is 1000.\\ 
$^\dagger$ We included WASP-43b for comparison as its atmospheric water abundance was recently measured from a transmission spectrum \citep{2014ApJ...793L..27K}.
}
\end{table*}

\subsection{The migration of short period, giant planets}
WASP-105b and WASP-107b could help us to understand the inward migration of hot Jupiters. 
To date, the orbits of planets in short orbits around cool stars (\teff\ < 6250 K) have been found to be near-circular and near-aligned with the stellar spins, whereas planets in longer orbits, therefore experiencing weaker tidal forces, tend to be eccentric and/or misaligned \citep{2012ApJ...757...18A, 2015ApJ...800L...9A}. 
This has been interpreted as evidence for high-eccentricity migration, in which a cold Jupiter is
perturbed into an eccentric, misaligned orbit that is then circularized, shortened, and realigned by tidal dissipation \citep{2012ApJ...757...18A}. 
Whilst high-eccentricity migration may be responsible for a substantial fraction of hot Jupiters, there is growing evidence that some migrated inwards to their current orbits via interaction with a protoplanetary disc. 
For example, WASP-84b is in a relatively wide orbit around a young star, indicating that it has experienced relatively weak tidal forcing over a short duration. Therefore, its near-circular and near-aligned orbit is suggestive of disc migration \citep{2014MNRAS.445.1114A, 2015ApJ...800L...9A}.
Further, WASP-47 comprises a hot Jupiter, a nearby super-Earth, a nearby Neptune, and a more distant Jupiter. The inner three planets are known to transit and the orbit of the hot Jupiter is near-aligned 
\citep{2012MNRAS.426..739H, 2015ApJ...812L..11S, 2015ApJ...812L..18B, 2016A&A...586A..93N}.
It seems likely that disc migration operated in both of these hot-Jupiter systems.

By measuring the spin-orbit angle for a sample of planets in relatively wide orbits, for which tidal effects will be smaller, we can determine the relative contributions of migration pathways. 
With scaled orbital separations of \arstar\ $\approx 18$, both WASP-105 and WASP-107 lie beyond the empirical boundary between aligned and misaligned systems (\arstar\ $\approx 15$) and so would be interesting targets in that respect. 
If the planets underwent high-eccentricity migration then we may expect their orbits to be eccentric and/or misaligned (e.g. \citealt{2007ApJ...669.1298F}), whereas near-circular and near-aligned orbits may be expected if they underwent planet-disc migration (e.g. \citealt{2009ApJ...705.1575M}). 
We placed an upper limit of $e < 0.04$ at 2 $\sigma$ for WASP-105b, but the constraint is much weaker for WASP-107b ($e < 0.4$ at 2 $\sigma$) due to the lower mass of the planet and therefore the smaller amplitude of the stellar reflex motion.
With a scaled separation of \arstar\ $\approx 9$, WASP-91b is expected to be in a near-aligned and near-circular orbit; indeed, we found $e < 0.07$ at 2\,$\sigma$.
The most common method employed to measure spin-orbit angles is to measure the apparent radial-velocity shift that occurs during transit (e.g. \citealt{2012ApJ...757...18A}).
The predicted semi-amplitude of the RV shift is 20, 13 and 35\,\ms\ for WASP-91, -105 and -107, respectively. 
An alternative possibility for WASP-107 is to infer the spin-orbit angle from the spot-crossing times measured over multiple transits \citep{2011ApJ...733..127S, 2011ApJ...740L..10N}. This could be done soon as WASP-107 was observed during Campaign 10 of the K2 mission (\citealt{2014PASP..126..398H}; Kepler Guest Observer proposal 8060, PI: Anderson).

\section*{Acknowledgements}
WASP-South is hosted by the South African Astronomical Observatory; we are grateful for their ongoing support and assistance. 
Funding for WASP comes from consortium universities and from the UK's Science and Technology Facilities Council. 
The Swiss {\it Euler} Telescope is operated by the University of Geneva, and is funded by the Swiss National Science Foundation. 
TRAPPIST is funded by the Belgian Fund for Scientific Research (Fond National de la Recherche Scientifique, FNRS) under the grant FRFC 2.5.594.09.F, with the participation of the Swiss National Science Fundation (SNF). 
M. Gillon and E. Jehin are FNRS Research Associates.
L. Delrez acknowledges support from the Gruber Foundation Fellowship.
This research has made use of the TEPCat catalogue of the physical properties of transiting planetary systems, which is maintained by John Southworth and is available at \url{http://www.astro.keele.ac.uk/jkt/tepcat}.

%\bibliography{dra}{}
%\bibliographystyle{aa}

\end{document}